# Research on Crop Water Status Monitoring and Diagnosis by Terahertz Imaging

**Bin Li[1],\*, Rong Wang[2] , Jianjun Ma[2],\*, Weihao Xu[1]**

[1]Beijing Research Center for Information Technology in Agriculture, Beijing, 100097 China

[2]School of Information and Electronics, Beijing Institute of Technology, Beijing, 100081 China

\* **Correspondence:**
Bin Li
lib@nercita.org.cn
Jianjun Ma
jianjun_ma@bit.edu.cn

**Abstract:** Leaf water content is regarded as an important index of crop health in agriculture. It is in great need to develop a rapid, non-destructive and accurate diagnose method which can measure and assess water level of different types of crops. In this research, winter wheat leaves were selected and the reliability and applicability of THz imaging technique for destructive and non-destructive water content monitoring were studied. Based on the measurement of THz transmission, smaller water loss in distal regions of a winter wheat leaf is shown when compared with that in basal regions during the natural dehydration process. A high correlation between the gravimetric water status level and THz transmission is observed during dehydration and after rehydration. The results show that the transmission of THz waves can be used to evaluate the water content of winter wheat leaves by destructive and non-destructive geometries, which suggests that THz imaging can be a good potential tool to measure the water content to diagnosis crop health in agriculture in a rapid, non-destructive and accurate way.

## 1    Introduction

Leaf water content of plants is usually regarded as an implication of their different exogenous and physiological conditions, which could affect productivity of crops [1]. So water content evaluation is regarded as an important indicator of crop health to farmers, agricultural scientists and plant physiologists. It provides essential information in irrigation management, assists to avoid irreversible damage, and thus substantially reduces or prevents yield losses [2]. Owing to the importance of leaf water content monitoring in agriculture, it is definitely necessary to develop a method which can measure and assess water level of different types of crops.

Present techniques in monitoing leaf water content can be seperated into two groups, destructive (gravimetric [3], pressure chamber [4]) and nondestructive  (chlorophyll fluorescence [5], visual estimate) techniques. Gravimetric measurement is usually considered as the most prominent destructive technique due to its high reliability and ease of handling. But it's not preferred for long-term researches of the same plant. Contactless measurement, as a kind of non-destructive technique, is always required over a long duration. But its accuracy limitation always exists and imprisons its applications.

THz imaging, firstly demonstrated 40 years ago [6], was found to be a promising technique for water content assessment with numerous advantages [7-9]. It can provide supplementary information which is not available by microwaves, infrared or optial images. Compared with images obtained lower frequencies, THz images own higher spatial resolution due to their shorter wavelength. Different from higher frequencies, a lot of common materials are almost transparent at THz range, including general packaging materials such as cardboard, plastic, paper and composites. Another advantage usually mentioned by researchers is that the THz radiation is non-ionizing [10], which makes it to be less harmful for biological tissues than other types of medical diagnostics.

THz imaging employs the relation between water content status and signal absorption of transmitted THz wave through the sample. It's high sensitivity to water content [10] makes it to be a sensitive and precise tool in evaluating plants [11, 12] and biotissue [13, 14] dehydration. So the development of THz imaging for highly accurate sensing of water status of biological tissues would help remarkably in such fields, such as food and other products quality control [15, 16], cancer diagnosis [17-21], and plant stress responses monitoring [7, 8, 22]. However, on the other side, this approach operating in the transmission geometry is also limited significantly due to such strong water absorption of the plants. So the reflection geometry was usually applied to test ticker biological samples [23, 24], which could help to reduce the effects of absorption due to water. But leaves are usually used for the testing in transmission geometry because of its thin thickness. There have been many researches concentrated on the water content assessment in leaves of plants, such as ginkgo [3], coffee [25], barley [8] and arabidopsis thaliana [26] by employing the transmission geometry as a general form. But more researches are still required for the water status monitoring for crops.

In this work, we investigate the applicability and reliability of THz imaging in testing water status in winter wheat leaves. A THz time domain spectroscopy (THz-TDS) system is employed in transmission geometry as we used previously in [27], due to the significant absorption by water [28] and the relatively small absorption by the dry-matter leaf [29]. Assessment on both the destructive and non-destructive methods is presented.

## 2     Experimental system and samples

### 2.1    Experimental measurements

The THz system used in this study is based on a standard commercial TERA K15 THz time-domain spectroscopy system (Menlo Systems Company, Germany) with a spectral range greater than 4 THz and frequency resolution less than 1.2 GHz. The schematic diagram of the whole system in transmission geometry can be seen in Fig. 1(a), which is more widely used as in [8, 29].

A Ti: Sapphire oscillator is employed to excite the THz pulse. It can produce pulses with a central wavelength of 1560nm with a 90 fs duration and 100 MHz repetition rate. The average output power can reach to 33mW. A collinear adapter is used to couple the transmitter and receiver to realize a collinear transmission transceiver. At the lower beginning frequency, a signal-to-noise ratio of better than 90dB can be achieved and that can be around 20dB at the high end frequency. The radiated THz beam can be focused to the samples with a minimum focal spot of 2 mm in diameter. A 20Hz scanning rate can be achieved with a minimum step size of 2.5 mm. An XY-stage is employ to form the image with the imaging area up to $5 \times 5$ cm$^2$. At the detection side, the sample substrate is positioned at the focal point of the detector for the measurement of reference signal in destruction method and air is tested as the reference in non-destruction method, followed by the sample signal detection. To isolate the influence of background noise and improve the stability of signal detection, we repeated twice for each measurement and averaged both to obtain reliable detection results.

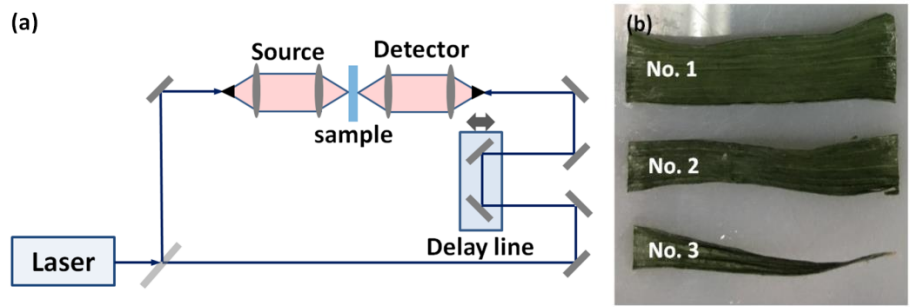

Figure 1 (a) Schematic diagram of the measurement setup; (b) photo of the samples (winter wheat leaves).

The 2-D image of the sample can be obtained by collecting, analyzing and reconstituting the transmitted THz spectral components. And each pixel of the image corresponds to the peak-to-peak amplitude of the transmitted THz pulse in time domain at one exact point of the leaf surface. It can indicate the obviously enhanced contrast difference [30]. However, in frequency domain , the image obtained based on dispersive coefficient, absorption coefficient, and/or other physical parameters of each pixel can usually yield a relatively better imaging [31]. And, due to the smaller wavelength with frequency increasing, the imaging resolution becomes better with the sacrifice of penetration depth because of the deceasing THz radiation and corresponding signal-to-noise ratio.

## 2.2 Sample preparation and configuration

Winter wheat plants (Jingdong 24) were cultivated under optimal illumination and watering conditions in Beijing Research Center for Information Technology in Agriculture. Three healthy leaves were cut from the plants. Each one were cut to three parts as shown in Fig. 1 (b) and measurements were taken on all three and then averaged. The fresh leaves were first used for measurement of THz transmission, and then weighed by an electronic balance for gravimetric water content evaluation. In every day of seven successive days, each leaf was tested on the identical condition and at the same time. All the leaves were kept in self-sealing bags after they were taken from plants and then placed in an incubator (4$^o$C) to decrease the water evaporation and tested immediately. Before the measurement, the leaves were wiped by sterile cotton, and then fixed onto the sample holder after the thickness was measured. To avoid scattering effect as much as possible, leaves with small surface roughness were chosen.

Together with the THz TDS measurement, the water content was tested gravimetrically as reference. The weight of sample leaves were first measured by weight scales with a precision of 0.1 mg at every experimental step. To isolate the influence of sample drying on the testing before and after THz measurements, the test was performed 2 times at each step. Then the average weight values can be obtained. At the last step when the leaf sample is dry, the value was considered as the weight of the dry component. The water content can be obtained as the following exression:

$$C_{m,\%} = \frac{(m - m_{dry})}{m} \times 100\% \qquad (1)$$

with m as the weight of the sample at the current experimental step, and $m_{dry}$ at the last measurement step.

## 3    Results and discussions

The first part of the measurements is the testing of the water content of the samples at several degrees of dryness by THz-TDS. During the experiments, the dryness of the samples can be created and controlled artificially. The sample was kept drying for several times with a time distance of 2 hours. So, different water status levels could be achieved within each experiment and then the sensitivity of the method could be evaluated. At the end, the sample was dried up as much as possible with without employing any other special instrument.

The THz time-domain waveforms were acquired for each pixel during in the imaging process. The peak-to-peak amplitude feature of the waveform under different humidity is plotted. During this measurement, a regular room temperature fluctuation around 23$^o$C was observed during the whole measurement period. Relative humidity was kept at 15%. The photos of the leaf samples by peak-to-peak amplitude values at each step are shown in Fig. 2. Even though there was only one sample for the test within each step, the testing was repeated 3 times to increase statistical significance. With the decreasing of the water content, the blue color disappears and becomes red. Meanwhile, the size of the leaf decreases because of the water loss inside. This indicates that the THz imaging can be used to recognize the water content of wheat leaf qualitatively.

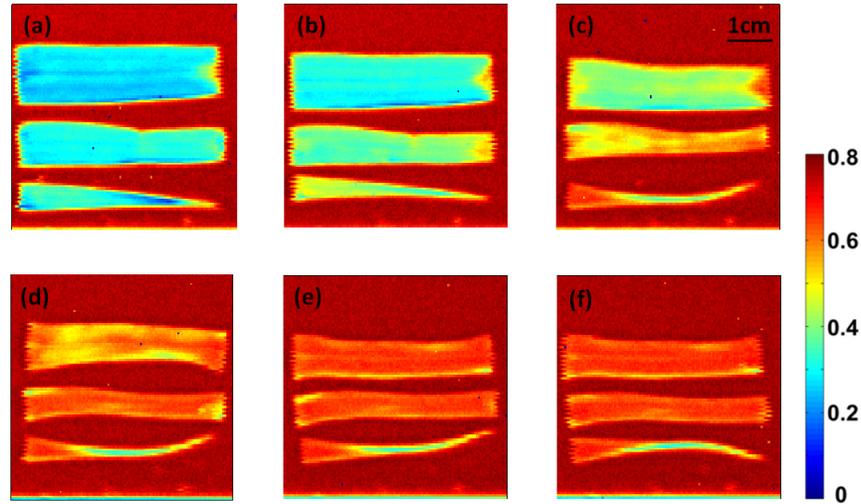

Figure 2 Time-domain imaging of wheat leaf in six different time points with a difference of 2 hours.

In addition, to quantitatively describe the spatial and temporal variability of the leaf water content, water status and peak-to-peak amplitude values at the same position on the leaf No.3 are plotted in Fig. 3. Here, the THz transmission in air is considered as a standard and the transmission without any samples is defined as 100%. During the entire leaf drying process of, the water content decreases in a nonlinear relationship with the increasing of the dehydration time. This is due to the intercellular spaces between the mesophyll cells, which do not differentiate palisade and spongy tissues [32, 33]. This spaces allow the passage of gases and makes the water loss rate to be slow in the later dehydration time [34]. Besides, the transmitted THz signal amplitude increses gradually along the basal, intermediate and distal regions at the same time, which implies that the water loss speed in the basal region is higher than others. This agrees well with the results in [35].

Here, we can also confirm that the THz transmission can be affected by both the water content and the measurement spot, which is investigated in [36]. Different spot corresponds to different leaf thickness/ composition. Repeated measurements of the wheat leaf at different spots could introduce a variation in transmission, even without a change in water content. So a successful monitor changes in plant water status should be done at the exact same spot with repeated measurements.

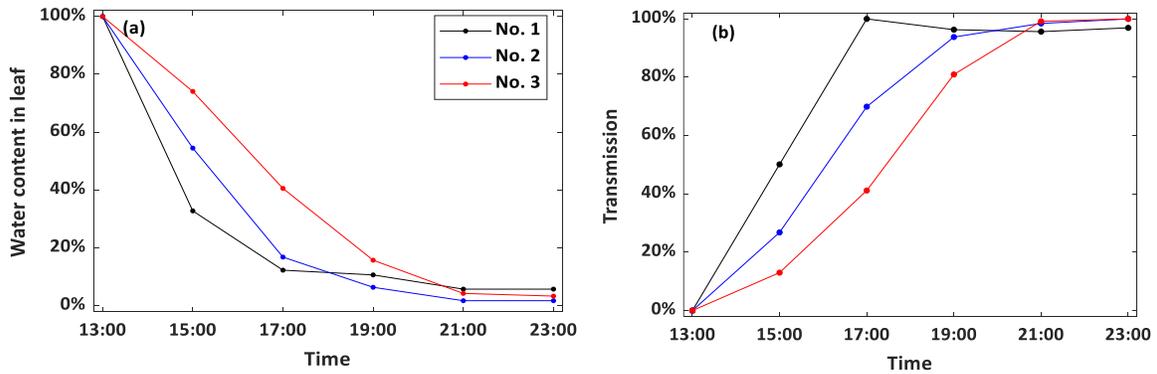

Figure 3 Variations of (a) water content and (b) THz image signals versus drying time.

Non-destructive technique is preferred for long-term leaf water content assessment, and can also provide vital information about vegetation productivity [37]. To test the reliability and applicability of this method, a piece of living wheat leaf was employed and the water content was measured at the same time (10:00am) on each day from May 16 to May 21, 2019 by the traditional gravimetric weighing method. The room temperature fluctuated around 23$^o$C regularly during the whole testing period with a 16$^o$C lowest temperature at night and a 25$^o$C maximum value during the day. Relative humidity fluctuated from 17% to 47% and showed no recognized pattern.The evolution of the water level inside the leaf is shown in Fig. 4(a). Fig. 4(b) shows the effects of it on the testing results at different frequencies from 0.2 – 1.0 THz. We conducted correlation analyses to confirm the consistency of results measured by different frequencies. From the Pearson analysis, the correlation coefficient between the water content status and THz signal transmission is calculated. The coefficient can be 0.9 at 0.6 THz, which indicates that the water content in the leaf can be monitored by the non-destructive THz imaging technique operating in transmission geometry.

The water content of soil and humidity of the lab should also affect the leaf water content. Here, we try to see whether we can predict the leaf water status by testing one of or both the parameters. The water content of soil was measured by the same gravimetric weighing method as above. The humidity was detected using a hygrometer. Fig. 4(c) and (d) show the variation of water content in wheat leaf with respect to water content in soil and relative humidity in the lab. We did this three times and find the same non-linear relationship, which implies that these two parameters cannot be used to predict the wheat leaf water content, even though they would affect the water status of the leaves.

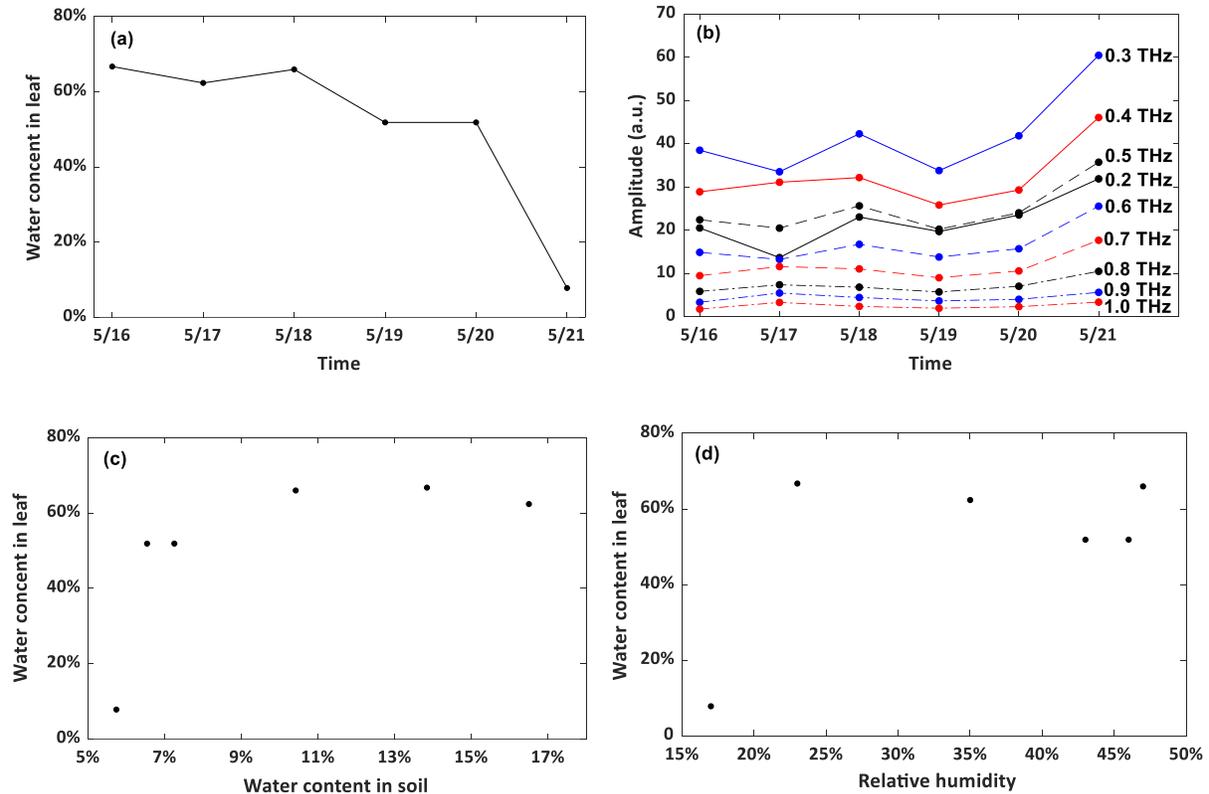

Figure 4 (a) water content evolution in the wheat leaf; (b) Frequency-domain THz imaging signal variation; (c) Correlation between water content in the leaf and soil; (d) Correlation between water content in the leaf and humidity in the lab room.

In above work, we did measurements during dehydration process to confirm that the THz imaging method can be used for water content assessment. A rehydration testing is also required to confirm our conclusion further. The plant was re-watered at 13:30 on May 21, 2019 and water content in the leaf was recorded at hourly intervals. Here, we measured the water content of the three leaves at the same height in the plant by the gravimetric weighing method and hope the results should be as close as possible to the leaf we monitored. Several different water content levels were obtained as shown in Fig. 5(a). The water content does not increase quickly after rehydration. But a recovery from hydration is still observed. A 0.8 correlation coefficient between the water content and imaging amplitude is obtained at 0.2 THz. This confirms that the THz imaging method allows for the detection of the leaf water content over time.

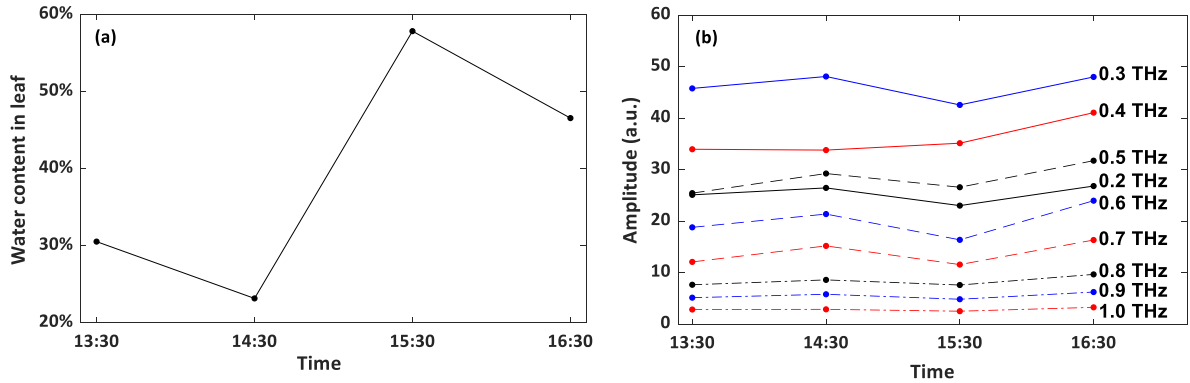

Figure 5  (a) Water content and (b) imaging signal after irrigation.

## 4    Conclusions

In this work, we demonstrate the application of THz imaging technique for the destructive and non-destructive water content sensing in winter wheat leaves. The relationship between the water content and the THz imaging is examined and a high correlation is observed during dehydration process and after rehydration. The water loss in distal, intermediate and basal regions is measured and higher water loss rate in the basal region is observed. This suggests that THz imaging has a good potential to measure the water content in different regions of wheat leaves in a simple and fast method. And we could confirm that repeated measurements at the same spot are highly accurate and reproducible. The leaf sample can be considered as a mixture of water and solid components[38] and a theoretical model considering surface roughness can be conducted to predict the signal transmission accurately, which will be our future directions.

**Funding:** We appreciate the support by Beijing Institute of Technology Research Fund Program for Young Scholars (No. 3050011181910), the Platform Construction Funded Program of Beijing Academy of Agriculture and Forestry Sciences (No. PT2020-29), National Key Research and Development Program of China (No. 2016YFD0702002) and Key-Area Research and Development Program of Guangdong Province (No. 2019B020219002).